\documentclass[aps,prd,twocolumn,nofootinbib,superscriptaddress]{revtex4}

\usepackage{latexsym}
\usepackage{amssymb}
\usepackage{epsfig,amsmath,graphics}

\newcommand{\gsim}{\lower.7ex\hbox{$\;\stackrel{\textstyle>}{\sim}\;$}}
\newcommand{\lsim}{\lower.7ex\hbox{$\;\stackrel{\textstyle<}{\sim}\;$}}
\newcommand{\OO}{\mathcal{O}}

\newcommand{\GeV}{\text{ GeV}}

\newcommand{\DO}{{\mbox{DO\hspace{-0.11in}$\not$\hspace{0.16in}}}}

\newcommand{\MET}{\mbox{$E_T\hspace{-0.23in}\not\hspace{0.18in}$}}
\newcommand{\SmMET}{E_T\hspace{-0.230in}\not\hspace{0.18in}}
\newcommand{\TiMET}{E_T\hspace{-0.155in}\not\hspace{0.14in}}

\begin{document}

\pagestyle{plain}

\title{
\begin{flushright}
\mbox{\normalsize SLAC-PUB-13149}
\end{flushright}
\vskip 15 pt

Searching for Directly Decaying Gluinos at the Tevatron}

\author{Johan Alwall, My-Phuong Le, Mariangela Lisanti and Jay G. Wacker}
\affiliation{
Theory Group, SLAC,  Menlo Park, CA 94025\\
Stanford Institute for Theoretical Physics, Stanford University,
Stanford, CA 94305 
}


\begin{abstract}
This letter describes how to perform searches over the complete kinematically-allowed parameter space for new pair-produced color octet particles that each subsequently decay into two jets plus missing energy at the Tevatron.  This letter shows that current searches can miss otherwise discoverable spectra of particles due to  CMSSM-motivated cuts.  Optimizing the $H_T$ and $\SmMET$ cuts expands the sensitivity of these searches.

\end{abstract}

 \maketitle

\section{Introduction}

In many theories beyond the Standard Model, there is a new color octet particle that decays into jets plus a stable neutral singlet.  This occurs, for example, in supersymmetry \cite{Dimopoulos:1981zb} and Universal Extra Dimensions \cite{Appelquist:2000nn}, as well as Randall-Sundrum \cite{Randall:1999ee} and Little Higgs models \cite{ArkaniHamed:2002qx}. 
As a result, jets plus missing transverse energy ($\MET$) is a promising experimental signature for new phenomena \cite{Alitti:1989ux, Portell:2006qb, Abazov:2007ww, CDF, MonojetSearches}.  

At present, the jets + $\MET$ searches at the Fermilab Tevatron are based upon the minimal supersymmetric standard model (MSSM) and look for production of gluinos ($\tilde{g}$) and squarks ($\tilde{q}$), the supersymmetric partners of gluons and quarks, respectively \cite{Abazov:2007ww, CDF}.  Both gluinos and squarks can decay to jets and a bino  ($\tilde{B}$), the supersymmetric partner of the photon.  The bino is stable, protected by a discrete R-parity, and is manifest as missing energy in the detector.  Different jet topologies are expected, depending on the relative masses of the gluinos and squarks.  

There are many parameters in the MSSM and setting mass bounds in a multidimensional parameter space is difficult.  This has lead to a great simplifying ans\"atz known as the CMSSM (or mSUGRA) parameterization of supersymmetry breaking \cite{CMSSM}.  This ans\"atz sets all the gaugino masses equal at the grand unified scale and runs them down to the weak scale, resulting in an approximately constant ratio between the gluino and bino masses ($m_{\tilde{g}}: m_{\tilde{B}} = 6:1$).  Thus, the mass ratio between the gluino and bino is never scanned when searching through CMSSM parameter space.  Since the bino is the LSP in most of the CMSSM parameter space, the restriction to unified gaugino masses means that there is a large region of kinematically-accessible gluinos where there are no known limits.  

The CMSSM parametrization is not representative of all supersymmetric models.  Other methods of supersymmetry breaking lead to different low-energy particle spectra. 
In anomaly mediation \cite{Randall:1998uk}, the wino can be the LSP; for instance, $m_{\tilde{g}}~:~m_{\tilde{W}}~\simeq~9:1$.  Mirage mediation \cite{Choi:2005ge}, in contrast, has nearly degenerate gauginos.  
A more comprehensive search strategy should be sensitive to all values of $m_{\tilde{g}}$ and $m_{\tilde{B}}$.  Currently, the tightest model-independent bound on gluinos is $51$ GeV and comes from thrust data at ALEPH and OPAL \cite{Kaplan:2008pt}.%
  
In this paper, we describe how bounds can be placed on all kinematically-allowed gluino and bino\footnote{Throughout this note, we will call the color octet a ``gluino'' and the neutral singlet the ``bino,'' though nothing more than the color and charge is denoted by these names.} masses.  We will treat the gluino as the first new colored particle and will assume that it only decays to the stable bino:
$\tilde{g} \rightarrow \bar{q}_1 \tilde{q}^*\rightarrow \bar{q}_1 q_2 \tilde{B}$.
The spin of the new color octet and singlet is not known a priori; the only selection rule we impose is that the two have the same statistics.  In practice, the spin dependence is a rescaling of the entire production cross section.  For our analysis, we will assume that the octet has spin 1/2, and will show how the results vary with cross section rescaling.  

We show how a set of optimized cuts for $\MET$ and $H_T = \sum_{\text{jets}} E_T$ can discover particles where the current Tevatron searches would not.   In order to show this, we model our searches on $\DO$'s searches for monojets \cite{MonojetSearches}, squarks and gluinos \cite{Abazov:2007ww}.  In keeping the searches closely tied to existing searches, we hope that our projected sensitivity is close to what is achievable and not swamped by unforeseen backgrounds.   


\section{Event Generation}
\label{Sec: Searches}

\subsection{Signal}

\begin{figure}[t] 
   \centering
   \includegraphics[width=3.24in]{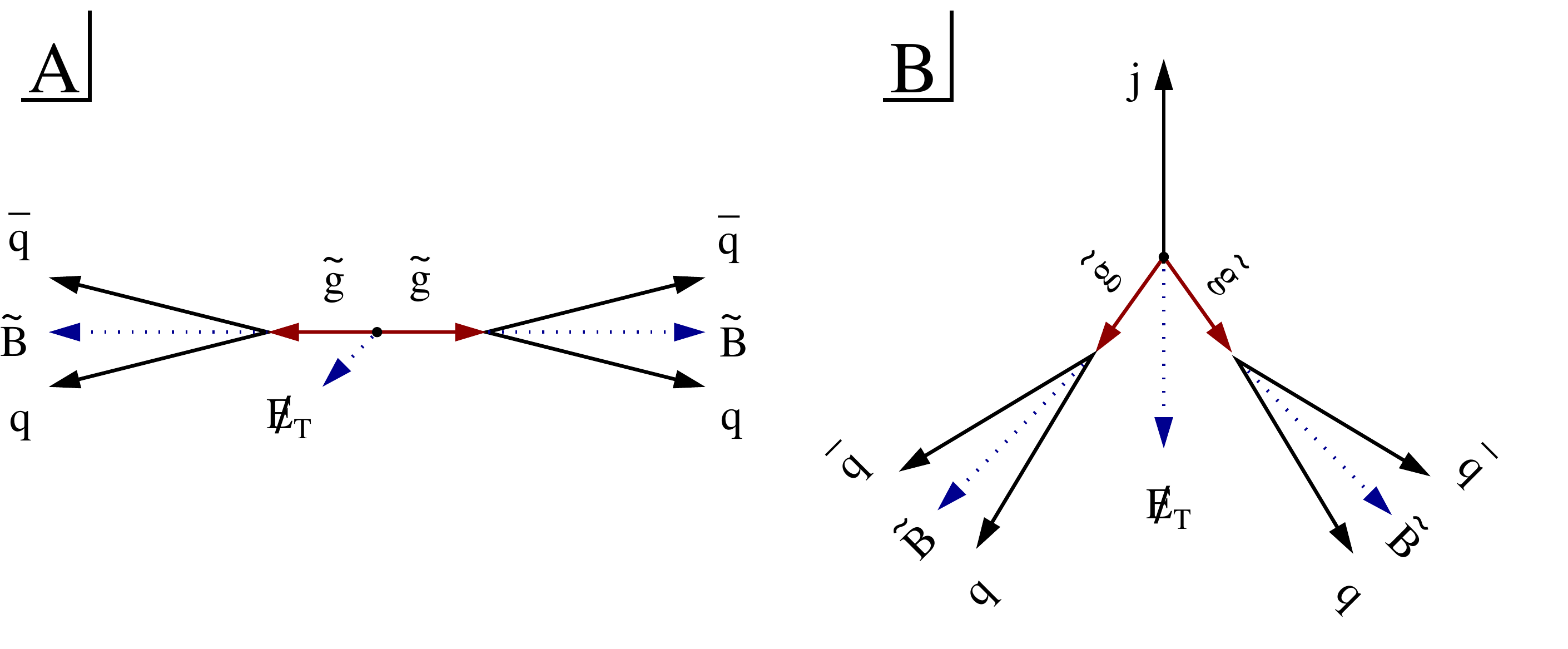}
   \caption{Boosted gluinos that are degenerate with the bino do not enhance the
   missing transverse energy when there is no hard initial- or final-state radiation.
   (A) illustrates the cancellation of the bino's $\MET$.  (B) shows how initial- or final-state radiation leads to a large amount of $\MET$ even if the gluino is degenerate with the bino.}
   \label{Fig: JetsFig}
\end{figure}

The number of jets expected as a result of gluino production at the Tevatron depends on the relative mass difference between the gluino and bino, $m_{\tilde{g}} - m_{\tilde{B}}$.  When the mass splitting is much larger than the bino mass, the search is not limited by phase space and  four or more well-separated jets are produced, as well as large missing transverse energy.
The situation is very different for light gluinos ($m_{\tilde{g}} \lesssim 200$ GeV) that are nearly degenerate with the bino.  Such light gluinos can be copiously produced at the Tevatron, with cross sections $\OO(10^2 \text{ pb})$, as compared to $\OO(10^{-2} \text{ pb})$ for their heavier counterparts ($m_{\tilde{g}} \gtrsim 400$ GeV).  Despite their large production cross sections, these events are challenging to detect because the jets from the decay are soft, with modest amounts of missing transverse energy.   Even if the gluinos are strongly boosted, the sum of the bino momenta will approximately cancel when reconstructing the missing transverse energy (Fig. \ref{Fig: JetsFig}A).
%
%
To discover a gluino degenerate with a bino, it is necessary to look at events where the gluino pair is boosted by the emission of hard QCD jets (Fig. \ref{Fig: JetsFig}B).  Therefore, initial-state radiation (ISR)  and final-state radiation (FSR) must be properly accounted for.

The correct inclusion of ISR/FSR with parton showering requires generating
gluino events with matrix elements.  We used MadGraph/MadEvent \cite{Alwall:2007st} 
to compute processes of the form
\begin{equation}
p \bar{p} \rightarrow \tilde{g} \tilde{g} + Nj,
\end{equation}
where $N=0, 1, 2$ is the multiplicity of QCD jets. The decay of the gluino
into a bino plus a quark and an antiquark, as well as parton showering
and hadronization of the final-state partons, was done in PYTHIA
6.4 \cite{Sjostrand:2006za}.

To ensure that no double counting of events occurs between the
matrix-element multi-parton events and the parton showers, a
version of the MLM matching procedure was used \cite{Alwall:2007fs}.
In this procedure, the matrix element multi-parton events and the
parton showers are constrained to occupy different kinematical
regions, separated using the $k_{\perp}$ jet measure:
\begin{eqnarray} 
\nonumber
&&d^2(i,j) = \Delta R^2_{ij} \min(p_{Ti}^2, p_{Tj}^2)\\
&&d^2(i,\text{beam}) = p_{Ti}^2,
\end{eqnarray}
where $\Delta R^2_{ij} = 2(\cosh\Delta\eta - \cos\Delta\phi)$ \cite{jetmeasure}. 
%
Matrix-element events are generated with some minimum cut-off $d(i, j)
= Q_{\text{min}}^{\text{ME}}$. After showering, the partons are
clustered into jets using the $k_T$ jet algorithm with a
$Q_{\text{min}}^{\text{PS}}>Q_\text{min}^\text{ME}$. The event is then
discarded unless all resulting jets are matched to partons in the
matrix-element event, $d(\text{parton},\text{jet}) <
Q_{\text{min}}^{\text{PS}}$. For events from the highest multiplicity
sample, extra jets softer than the softest matrix-element parton are
allowed.  This procedure avoids double-counting jets, and results in
continuous and smooth differential distributions for all jet
observables.

The matching parameters ($Q_{\text{min}}^{\text{ME}} $ and
$Q_{\text{min}}^{\text{PS}}$) should be chosen resonably far below the
factorization scale of the process.  For gluino production, the
parameters were:
\begin{equation}
Q_{\text{min}}^{\text{ME}} = 20 \mbox{ GeV and  } Q_{\text{min}}^{\text{PS}}= 30 \mbox{ GeV.}    
\end{equation}

The simulations were done using the CTEQ6L1 PDF
\cite{Pumplin:2002vw} and with the renormalization and factorization scales
set to the gluino mass.  The cross sections
were rescaled to the next-to-leading-order (NLO) cross sections
obtained using Prospino 2.0 \cite{Beenakker:1996ch}.

Finally, we used PGS \cite{PGS} for detector simulation, with a cone
jet algorithm with $\Delta R = 0.5$.  As a check on this procedure, we
compared our results to the signal point given in \cite{Abazov:2007ww}
and found that they agreed to within 10\%.

\subsection{Backgrounds}
\label{Sec: Background}

The three dominant Standard Model backgrounds that contribute to the jets plus missing energy searches are: $W^\pm/Z^0$ + jets, $t \bar{t}$, and QCD.  There are several smaller sources of missing energy that include single top and di-boson production, but these make up a very small fraction of the background and are not included in this study.

The $W^\pm/Z^0 + n j$ and $t \bar{t}$ backgrounds were generated using MadGraph/MadEvent and then showered and hadronized using PYTHIA.  PGS was used to reconstruct the jets.  MLM matching was applied up to three jets for the $W^\pm/Z^0$ background, with the parameters $Q_{\text{min}}^{\text{ME}} = 10$ GeV and $Q_{\text{min}}^{\text{PS}} = 15$ GeV.  The top background was matched up to two jets with $Q_{\text{min}}^{\text{ME}} = 14$ GeV and $Q_{\text{min}}^{\text{PS}} = 20$ GeV.  Events containing isolated leptons with $p_T \geq$ 10 GeV were vetoed to reduce background contributions from leptonically decaying $W^\pm$ bosons.  To reject cases of $\MET$ from jet energy mismeasurement, a lower bound of $90^{\circ}$ and $50^{\circ}$ was placed on the azimuthal angle between $\MET$ and the first and second hardest jets, respectively.  An acoplanarity cut of $<165^\circ$ was applied to the two hardest jets.  Because the $\DO$ analysis did not veto hadronically decaying tau leptons, all taus were treated as jets in this study.  

Simulation of the missing energy background from QCD is beyond the scope of PYTHIA and PGS, and was therefore not done in this work.  However, to avoid the regions where jet and calorimeter mismeasurements become the dominant background, a lower limit of $\MET > 100$ GeV was imposed.  Additionally, in the dijet analysis, the azimuthal angle between the $\MET$ and any jet with $p_T \geq 15$ GeV and $| \eta | \leq$ 2.5 was bounded from below by $40^{\circ}$.  This cut was not placed on the threejet or multijet samples because of the large jet multiplicities in these cases.


For each of the $W^\pm/Z^0+nj$ and $t\bar{t}$ backgrounds, 500K events were generated.  The results reproduce the shape and scale of the $\MET$ and $H_T$ distributions published by the $\DO$ collaboration in \cite{Abazov:2007ww} for 1fb$^{-1}$.  For the dijet case, where the most statistics are available, the correspondence with the $\DO$ result is $\pm 20\%$.  With the threejet and multijet cuts, the result for the $t \bar{t}$ background is similar, while the $W^\pm/Z^0 + nj$ backgrounds reproduce the $\DO$ result to within $30 - 40 \%$ for the threejet and multijet cases.  The increased uncertainty may result from insufficient statistics to fully populate the tails of the $\MET$ and $H_T$ distributions.  The PGS probability of losing a lepton may also contribute to the relative uncertainties for the $W^\pm+nj$ background.   Heavy flavor jet contributions were found to contribute $2\%$ to the $W^\pm/Z^0$ backgrounds, which is well below the uncertainties that arise from not having NLO calculations for these processes and from using PGS.  

\section{Projected Reach of Searches}
\label{Sec:Reach}
A gluino search should have broad acceptances over a wide range of kinematical parameter space; it should be sensitive to cases where the gluino and bino are nearly degenerate, as well as cases where the gluino is far heavier than the bino.  As already discussed, the number of jets and $\MET$ depend strongly on the mass differerence between the gluino and bino.  
Because the signal changes dramatically as the masses of the gluino and bino are varied, it is necessary to design searches  that are general, but not closely tied to the kinematics.  We divided events into four mutually exclusive searches for $\MET$ plus $1j$, $2j$, $3j$ and $4^+j$, respectively.  For convenience, we keep the $nj+\MET$ classification fixed for all gluino and bino masses (see Table \ref{Tab: Cuts}).    
These selection criteria were modeled after those used in $\DO$'s existing search \mbox{\cite{Abazov:2007ww}. \!\!\!}\footnote{It should be noted, however, that the $\DO$ searches are inclusive because each is designed to look for separate gluino/squark production modes (i.e., $pp \rightarrow$ $\tilde{q} \tilde{q}$, $\tilde{q}\tilde{g}$, $\tilde{g} \tilde{g}$). }   These exclusive searches can be statistically combined to provide stronger constraints.     
 \begin{table}
\begin{center}
\begin{tabular}{|c||c|c|c|c|}
\hline
&$1j+\SmMET$&$2j+\SmMET$&$3j+\SmMET$&$4^+j+\SmMET$\\
\hline\hline
$E_{T\,j_1}$&$\ge 150$&$\ge 35$&$\ge 35$&$\ge 35$\\
$E_{T\,j_2}$&$< 35$&$\ge 35$&$\ge 35$&$\ge 35$\\
$E_{T\,j_3}$&$< 35$&$<35$&$\ge 35$&$\ge 35$\\
$E_{T\,j_4}$&$< 20$&$<20$&$<20$&$\ge 20$\\
\hline
\end{tabular}
\caption{\label{Tab: Cuts} Summary of the selection criteria for
the four non-overlapping searches.  The two hardest jets are required to be central ($| \eta | \leq 0.8$).  All other jets must have $| \eta | \leq 2.5$.}
\end{center}
\end{table}

\begin{figure}[t] 
   \centering
   \includegraphics[width=3.35in]{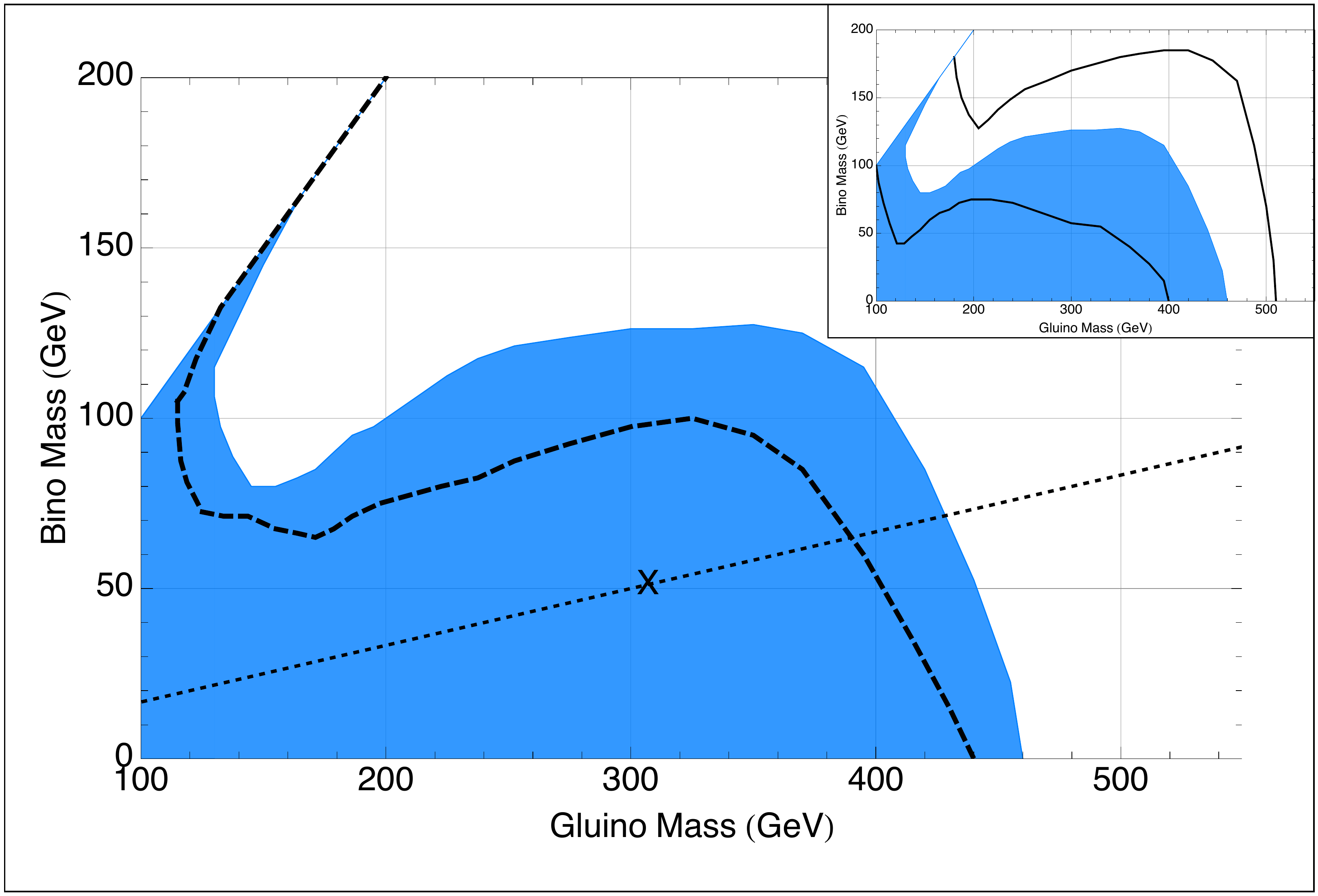}
   \caption{ The 95\% gluino-bino exclusion curve for $\DO$ at 4 fb$^{-1}$ for $S/B >1$.  The dashed line shows the corresponding exclusion region using $\DO$'s non-optimized cuts.  The masses allowed in the CMSSM are represented by the dotted line; the ``X'' marks the current $\DO$ limit on the gluino mass at 2.1 fb$^{-1}$ (see text for details) \cite{Abazov:2007ww}. The inset shows the effect of scaling the production cross section for the case of $S/B>1$.  The solid lines show the exclusion region for $\sigma/3$ (bottom) and $3\sigma$ (top).}
   \label{fig:exclusion}
\end{figure}

Two cuts are placed on each search: $H_T^{\text{min}}$ and $\MET^{\text{min}}$.  In the $\DO$ analysis, the $H_T$ and $\MET$ cuts are constant for each search.  The signal (as a function of the gluino and bino masses) and Standard Model background are very sensitive to these cuts.
 To maximize the discovery potential, these two cuts should be optimized for all gluino and bino masses.  For a given gluino and bino mass,  the significance  ($S/\sqrt{S+B}$) is maximized over $H_T^{\text{min}}$ and $\MET^{\text{min}}$ in each $nj+\MET$ search.  Due to the uncertainty in the background calculations, the $S/B$ was not allowed to drop beneath the conservative limit of $S/B > 1$.  More aggressive bounds on $S/B$ may also be considered; $\DO$, for instance, claims a systematic uncertainty of $\OO(30\%)$ in their background measurements \cite{Abazov:2007ww}.  The resulting 95\% sensitivity plot using the optimized $H_T$ and $\MET$ cuts is shown in Fig. \ref{fig:exclusion}.  The corresponding inset illustrates the effect of varying the production cross section.

For light and degenerate gluinos, the $1j+\MET$ and $2j+\MET$ searches both have good sensitivity.  In an intermediate region, the $2j+\MET$, $3j+\MET$ and $4^+j+\MET$ all cover with some success, but there appears to be a coverage gap where no search does particularly well.  If one does not impose a $S/B$ requirement, a lot of the gap can be covered, but background calculations are probably not sufficiently precise to probe small $S/B$.
For massive, non-degenerate gluinos, the $3j+\MET$ and $4^+j+\MET$ both give good sensitivity, with the $4^+j+\MET$ giving slightly larger statistical significance.

In the exclusion plot, the $\MET$ and $H_T$ cuts were optimized for each point in gluino-bino parameter space.  However, for gluino masses $200 \GeV \lsim m_{\tilde{g}} \lsim 350\GeV$, where the monojet search gives no contribution, we found that the exclusion region does not markedly change if the following set of generic cuts are placed:
\begin{eqnarray}
\nonumber
\!\!\!\!\!\!(H_T,\MET) &\ge& (150, 100)_{2j+\TiMET},\\
&& (150, 100)_{3j+\TiMET}, (200, 100)_{4^+j+\TiMET}.
\end{eqnarray}
As a comparison, the cuts used in the $\DO$ analysis are 
\begin{eqnarray}
\nonumber
\!\!\!\!\!\!\!(H_T,\MET) &\ge& (325, 225)_{2j+\TiMET},\\
&& (375, 175)_{3j+\TiMET}, (400, 100)_{4^+j+\TiMET}.
\end{eqnarray}
The lowered cuts provide better coverage for intermediate mass gluinos, as indicated in Fig. \ref{fig:exclusion}.  For $m_{\tilde{g}} \lesssim 200$ GeV, we place tighter cuts on the monojet and dijet samples than $\DO$ does.  While $\DO$ technically has statistical significance in this region with their existing cuts, their signal-to-background ratio is low.  Because of the admitted difficulties in calculating the Standard Model backgrounds, setting exclusions with a low signal-to-background should not be done and fortunately can be avoided by tightening the $H_T$ and $\MET$ cuts.  Similarly, for larger gluino masses, the generic cuts are no longer effective and it is necessary to use the optimized cuts, which are tighter than \DO's.  

\section{Conclusions and Outlook}

In this paper, we describe the sensitivity that $\DO$ has in searching for gluinos away from the CMSSM hypothesis in jets + $\MET$ searches.  It was assumed that the gluino only decayed to two jets and a stable bino.  However, many variants of this decay are possible and the search presented here can be generalized accordingly.  

One might, for example, consider the case where the gluino decays dominantly to bottom quarks and heavy flavor tagging can be used advantageously.  Cascade decays are another important possibility.  Decay chains have a significant effect upon the searches because they convert missing energy into visible energy.  In this case, additional parameters, such as  the intermediate particle masses and the relevant branching ratios, must be considered.  In the CMSSM, the branching
ratio of the gluino into the wino is roughly 80\%.  This is the dominant decay affecting the $\DO$ gluino mass bound in CMSSM parameter space (see Fig. 2).  While this cascade decay may be representative of many models that have gluino-like objects, the fixed mass ratio and branching ratio
are again artifacts of the CMSSM.   A more thorough examination of cascade decays should be considered.

In addition to alternate decay routes for the gluino, alternate production modes are important when there are additional particles that are kinematically accessible.
In this paper, it was assumed that the squarks are kinematically inaccessible at the Tevatron;  however, if the squarks are accessible, $\tilde{g} \tilde{q}$ and $\tilde{q} \tilde{q}$ production channels
could lead to additional discovery possibilities.  For instance, a gluino that is degenerate with the bino could be produced with a significantly heavier squark.  The squark's subsequent cascade decay to the bino will produce a great deal of visible energy in the event and may be more visible than gluino pair production.\footnote{We thank M. Ibe and R. Harnik for this observation.}  

Finally, in the degenerate gluino region, it may be beneficial to use a mono-photon search rather than a monojet search \cite{Abazov:2008kp}.\footnote{We thank F. Petriello for pointing this out}  Preliminary estimates of the reach of the mono-photon search show that it is not as effective as the monojet search.  This is likely due to the absence of final-state photon radiation from the gluinos.  However, it may be possible to better optimize the mono-photon search, because the Standard Model backgrounds are easier to understand in this case.    

Ultimately, a model-independent search for jets plus missing energy would be ideal.  We believe that our exclusive $n j+ \MET$ searches, with results presented in an exclusion plot as a function of $H_T$ and $\MET$, would provide significant coverage for these alternate channels \cite{LongPaper}.   This analysis should be carried forward to the LHC to ensure that the searches discover all possible supersymmetric spectra.  The general philosophy of parameterizing the kinematics of the decay can be easily carried over.   The main changes are in redefining the $H_T$ and $\MET$ cuts, as well as the hard jet energy scale.  We expect a similar shape to the sensitivity curve seen in Fig. \ref{fig:exclusion}, but at higher values for the gluino and bino masses.  Therefore, it is unlikely that there will be a gap in gluino-bino masses where neither the Tevatron nor the LHC  has sensitivity.  

\section*{Acknowledgements}
We would like to thank  Jean-Francois Grivaz, Andy Haas, Roni Harnik, Masahiro Ibe, Greg Landsberg, Frank Petriello, and Patrice Verdier for helpful discussions.   JA, M-PL, ML, and JGW are supported by the DOE under contract DE-AC03-76SF00515 and partially by the NSF under grant PHY-0244728.  JA is supported by the Swedish Research Council.  ML is supported by NDSEG and Soros fellowships.

\providecommand{\href}[2]{#2}\begingroup\raggedright

\endgroup

\end{document}